\documentclass[aps,pra,onecolumn,superscriptaddress,showpacs]{revtex4-1}

\usepackage{amssymb,amsmath}
\usepackage{graphicx}
\DeclareGraphicsExtensions{.png}
\usepackage[sort&compress]{natbib}
\usepackage{color}
\usepackage{multirow}
\usepackage{sidecap}
\usepackage{graphicx}
\usepackage{dcolumn}% Align table columns on decimal point
\usepackage{bm}% bold math

\def \co {{CoV$_2$O$_4$} }
\def \mn {{MnV$_2$O$_4$} }

\begin{document}
\title{Magnetic Frustration Driven by Itinerancy in Spinel \co}
\author{J. H. Lee}
\altaffiliation{jjjun97@gmail.com}
\affiliation{School of Energy and Chemical Engineering, Ulsan National Institute of Science and Technology, Ulsan 44919, Republic of Korea}
\affiliation{Materials Science and Technology Division, Oak Ridge National Laboratory, Oak Ridge, Tennessee, 37831, USA}
\author{J. Ma}
\affiliation{Key Laboratory of Artificial Structures and Quantum Control,Department of Physics and Astronomy, Shanghai Jiao Tong University, Shanghai 200240, China}
\affiliation{Collaborative Innovation Center of Advanced Microstructures, Nanjing, Jiangsu 210093, People's Republic of China}
\affiliation{Department of Physics and Astronomy, University of Tennessee, Knoxville, Tennessee 37996, USA}
\author{S. E. Hahn}
\affiliation{Neutron Data Analysis and Visualization Division, Oak Ridge National Laboratory, Oak Ridge, Tennessee 37831, USA}
\author{H. B. Cao}
\affiliation{Quantum Condensed Matter Division, Oak Ridge National Laboratory, Oak Ridge, Tennessee 37831, USA}
\author{Tao Hong}
\affiliation{Quantum Condensed Matter Division, Oak Ridge National Laboratory, Oak Ridge, Tennessee 37831, USA}
\author{S. Okamoto}
\affiliation{Materials Science and Technology Division, Oak Ridge National Laboratory, Oak Ridge, Tennessee, 37831, USA}
\author{H. D. Zhou}
\affiliation{Department of Physics and Astronomy, University of Tennessee, Knoxville, Tennessee 37996, USA}
\author{M. Matsuda}
\affiliation{Quantum Condensed Matter Division, Oak Ridge National Laboratory, Oak Ridge, Tennessee 37831, USA}
\author{R. S. Fishman}
\affiliation{Materials Science and Technology Division, Oak Ridge National Laboratory, Oak Ridge, Tennessee, 37831, USA}
%\date{\today}
% ($A$V$_2$O$_4$, $A$=Mg,Zn,Cd,Fe,Mn)
\begin{abstract}
Localized spins and itinerant electrons rarely coexist in geometrically-frustrated spinel lattices. We show that the spinel CoV$_2$O$_4$ stands at the crossover from insulating to itinerant behavior and exhibits a complex interplay between localized spins and itinerant electrons. In contrast to the expected paramagnetism, localized spins supported by enhanced exchange couplings are frustrated by the effects of delocalized electrons. This frustration produces a non-collinear spin state and may be responsible for macroscopic spin-glass behavior. Competing phases can be uncovered by external perturbations such as pressure or magnetic field, which enhance the frustration.

\end{abstract}


\maketitle

The interplay between localized-spin and itinerant-electron behavior
in geometrically-frustrated systems \cite{frust} is responsible for
many intriguing phenomena such as a spin-liquid state \cite{Nakatsuji06}, heavy-fermion behavior \cite{Lee13},
spin-ice conduction \cite{Udagawa12}, and exotic phases \cite{Hanasaki07,Iguchi09,Kumar10}.
While the interplay between the localized spins and itinerant electrons
has been investigated intensively on pyrochlores $A_2B_2$O$_7$ where both $A$ and $B$ sublattices
are frustrated \cite{Nakatsuji06,Lee13,Udagawa12,Hanasaki07,Iguchi09},
it has rarely been explored on spinel systems $AB_2$O$_4$, where only the sublattice $B$ is frustrated.
In the spinels, the electronic itinerancy on sublattice $B$ can be controlled by chemical pressure on sublattice $A$, which
enhances magnetic frustration.
On the other hand, magnetic $A$ site ions can suppress the frustration through their
magnetic interactions with the localized spins on sublattice $B$.
Thus, one can anticipate a rich interplay between localized spins and itinerant electrons.

In the spinel vanadates $A$V$_2$O$_4$, the chemical pressure exerted by a
small $A$-site cation can introduce itinerancy. Since it has the smallest magnetic $A$-site cation
of any known spinel vanadate, CoV$_2$O$_4$
would be the ideal system to study the interplay between itinerancy and localized spins.
In most spinels vanadates ($A$V$_2$O$_4$, $A$=Mn, Fe, Cd, Zn, Mg), non-collinear (NC) spin states
are produced by the orbital ordering (OO) of partially-filled $d$-electrons on the V site, which
relieves geometric frustration through a tetragonal distortion \cite{Kismarahardja1, Nishiguchi, Lee, Wheeler, Garlea, Katsufuji, MacDougall1}.
However, even in a cubic phase without OO, CoV$_2$O$_4$ macroscopically exhibits NC and glassy spin states \cite{Kiswandhi,Huang} as well
as other magnetic anomalies \cite{Kismarahardja}.
Therefore, the NC spin states in CoV$_2$O$_4$ demand a detailed study.

Although the macroscopic spin-glass behavior of Mn$_{1-x}$Co$_x$V$_2$O$_4$ is enhanced by Co-doping \cite{Kiswandhi},
CoV$_2$O$_4$ has higher magnetic ordering temperatures
than compounds Mn$_{1-x}$Co$_x$V$_2$O$_4$ with $x<1$.
Indeed, CoV$_2$O$_4$ has higher collinear ($T_{\rm CL}$) and NC ($T_{\rm NC}$) spin transition temperatures
than any other spinel.  This stands in marked contrast to the
pyrochlores \cite{Hanasaki07,Iguchi09}, where spin-glass phases have lower ordering temperatures.

While its earlier characterization was hampered by the difficulty of fabricating single crystals,
recent experiments on single crystals of \co have reported anomalous physical
and magnetic properties \cite{Kiswandhi,Kismarahardja,Kismarahardja1,Huang}.
This paper clarifies the origin of the NC spin states of CoV$_2$O$_4$ by using density functional theory (DFT)
and spin models to interpret
neutron-scattering measurements on CoV$_2$O$_4$ single crystals.

In contrast to previous macroscopic measurements \cite{Huang,Kiswandhi},
our neutron scattering measurements show CoV$_2$O$_4$ to be an ordered magnet rather than a spin glass.
Nevertheless, some latent factors of \co can enhance frustration and drive it into a spin glass
with the help of external perturbations.
Chemically-driven pressure by Co increases itinerancy in CoV$_2$O$_4$.
This itinerancy weakens the OO and thus enhances magnetic and structural isotropies.
The frustration fostered by that isotropy \cite{Ising}
may be responsible for macroscopic glassy behavior in a magnetic field \cite{Huang,Kiswandhi}.
Due to the enhanced frustration, external perturbation such as pressure or magnetic field could uncover
novel magnetic phases in cubic CoV$_2$O$_4$.

\vspace{4mm}
\noindent{\bf \large Results}

\noindent{\bf NC spin states in cubic phase \co}

\begin{figure}

\includegraphics[width=98mm,   trim=3mm 10mm 0mm 5mm]{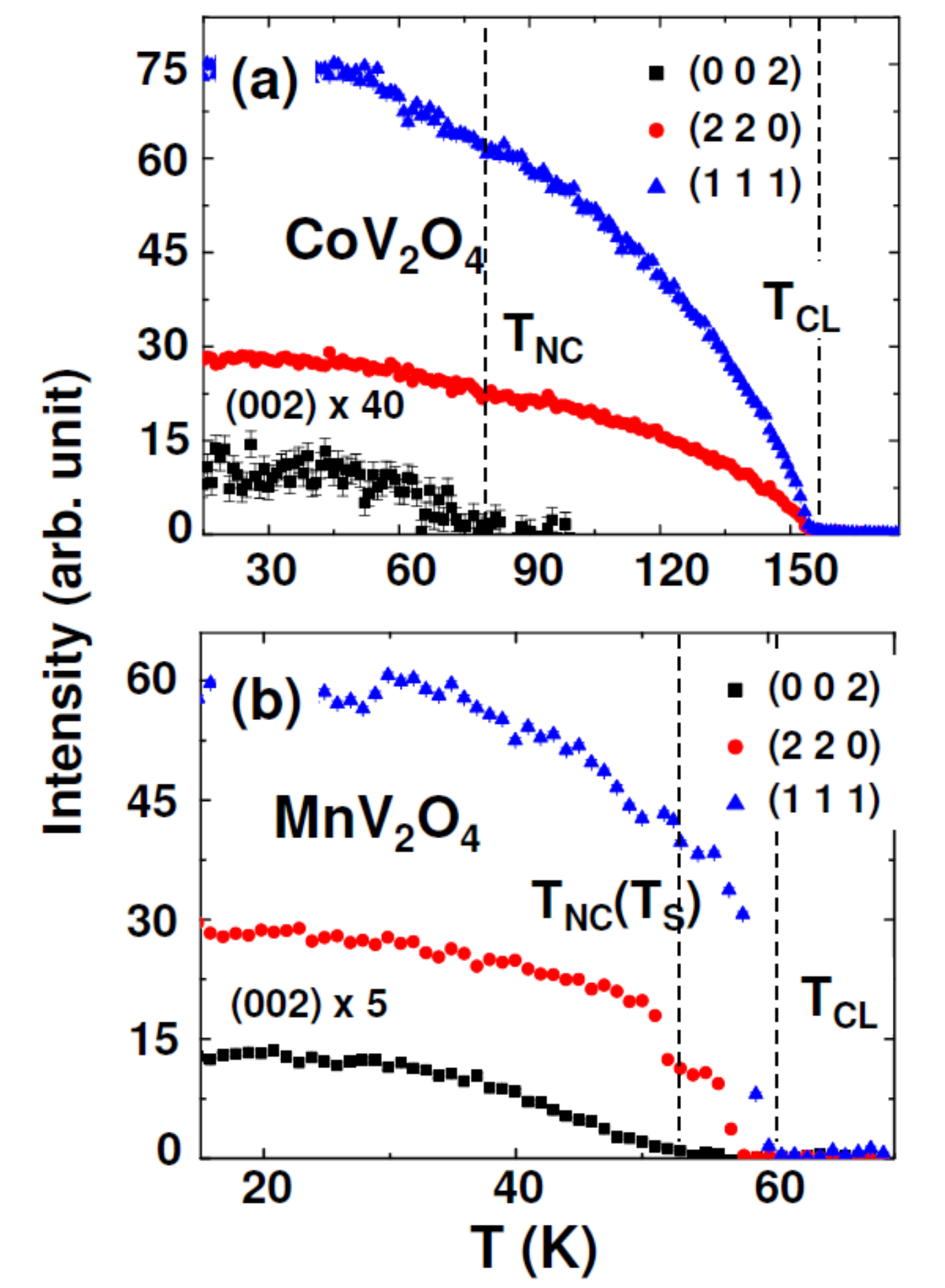}\\
\caption{\label{Bragg}\textbf{NC spin states in cubic CoV$_2$O$_4$ compared to tetragonal MnV$_2$O$_4$.}
Temperature dependence of the (111) (triangles), (220) (circles), and (002) (squares)
Bragg peak intensities for CoV$_2$O$_4$ \textbf {(a)}  and MnV$_2$O$_4$ \textbf {(b)}
measured by neutron diffraction at HB-3A.
The peak intensities of (111) and (220) above the magnetic transition temperature are fully
from the nuclear structure and were subtracted.
The (002) peak is not allowed from the structural symmetry and fully originated from the magnetic scattering.
The background was subtracted.
All the magnetic peaks observed by our neutron diffraction are instrument resolution limited
besides the peak broadening caused by the structural transition for MnV$_2$O$_4$,
and thus indicate the long range ordered magnetic moments.
The strongly-reduced intensity of (002) peak in CoV$_2$O$_4$ indicates that only tiny amount of V spin orders,
which is caused by enhanced itinerancy.}
\end{figure}

Figure~\ref{Bragg} compares the temperature dependence of the (002), (220), and (111)
Bragg peaks in cubic CoV$_2$O$_4$ and tetragonal MnV$_2$O$_4$.
The temperature dependence of the two end-compounds was recently compared
to that of the intermediates compounds Mn$_{1-x}$Co$_x$V$_2$O$_4$ (x = 0.2, 0.4, 0.6, 0.8) \cite{Jie}.
At the symmetry-allowed Bragg positions of (220) and (111), a ferrimagnetic (FIM) signals occur below $T_{CL}$.
While the (002) peak is forbidden by structural symmetry, the observed scattering below $T_{NC}$
indicates the formation of an antiferromagnetic (AFM) component in the $ab$-plane.

The (002) magnetic reflection indicates that the collinear (CL) to NC
magnetic transition at $T_{NC}$ coincides with the structural transition at $T_S$ in MnV$_2$O$_4$ \cite{Garlea}.
The (220) and (111) Bragg peaks show the FIM CL spin transition ($T_{CL}$$\sim$150 K)
and confirm recent X-ray diffraction and heat capacity measurements~\cite{Kiswandhi} that
indicate the disappearance of the structural transition in CoV$_2$O$_4$.
However, the additional magnetic transition in CoV$_2$O$_4$
indicated by the (002) peak has an enhanced transition temperature $T_{NC}$$\sim$75 K compared to $T_{NC}$$\sim$57 K
in MnV$_2$O$_4$.  Despite the induced itinerancy, this (002) peak indicates the
formation of the two-in/two-out (TI/TO) spin configuration on the V-sublattice.
Based on the diffraction data, this spin configuration
is similar to the spin configurations in MnV$_2$O$_4$ \cite{Magee} and FeV$_2$O$_4$ \cite{MacDougall}.
As shown in Supplement, the full widths at half maximum of the Bragg peaks (111) and (220) are not broadened
below $T_{NC}$.  Consequently, \co is an ordered magnet and not yet a spin glass.

Although \co is not a spin glass, magnetic frustration is still induced by electronic itinerancy.
The ordered magnetic moment refined from the peaks in Fig.~\ref{Bragg} is 0.47(3)$\mu_B$/V in \co
which is significantly reduced from 0.95(4)$\mu_B$/V in MnV$_2$O$_4$ due to the increased itinerancy.
While the moment in \co is small, the paramagnetic metallic state was expected down to zero temperature
when the inter-vanadium distance $R_{\rm V-V}$=2.97 \AA~lies below the critical value (2.98 \AA) \cite{Canosa}.
Since the TI/TO state originates from OO in tetragonal compounds \cite{Magee,MacDougall},
the isosymmetric TI/TO state in cubic \co without any OO
must have a different origin associated with its itinerancy and frustration.

\vspace{4mm}
\noindent{\bf Single-ion anisotropy suppressed by itinerancy}

\begin{figure}
\includegraphics[width=0.67\textwidth]{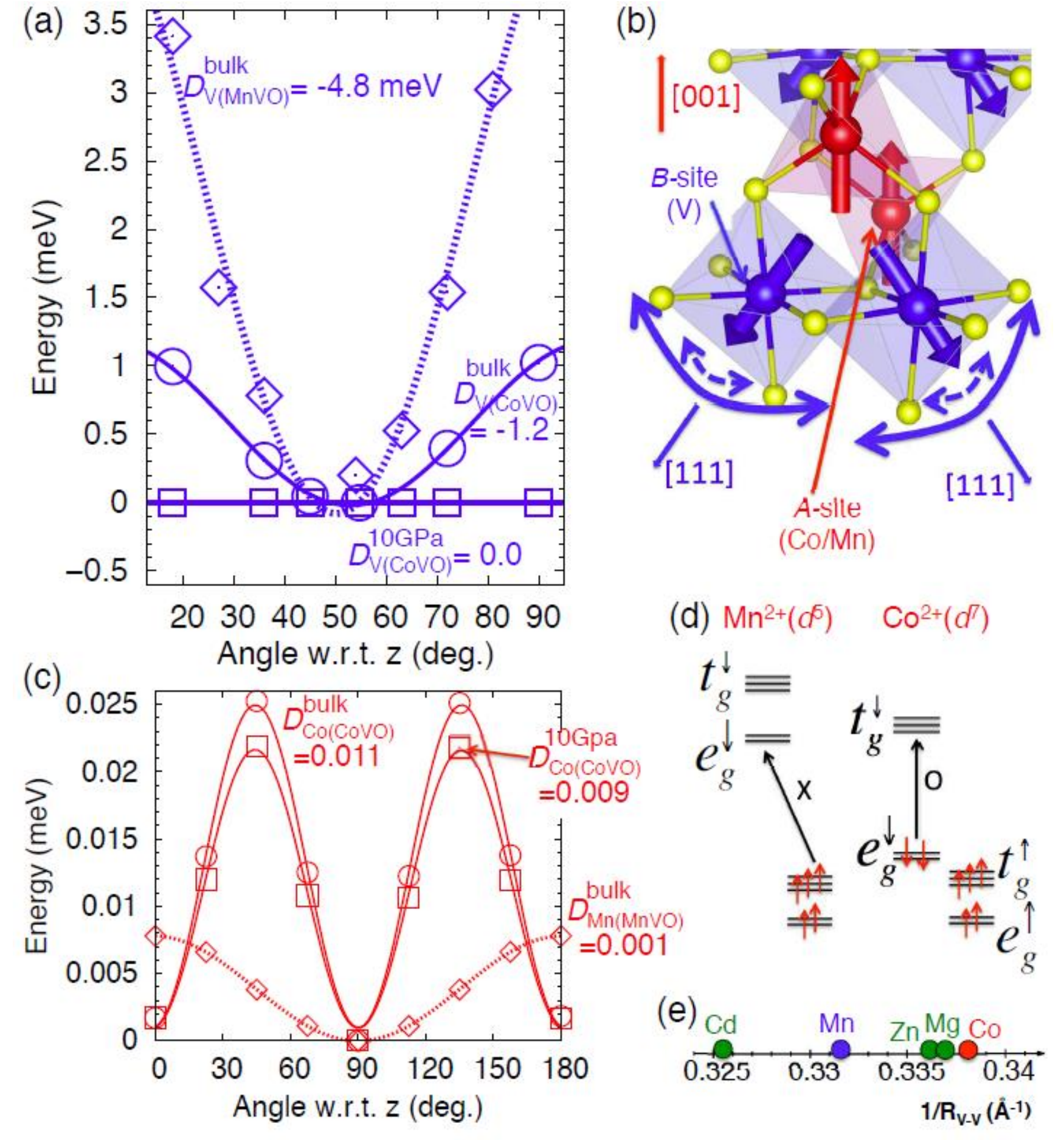}\\
\caption{\label{D}
  \textbf{ Reduced single-ion anisotropy (SIA) of V in CoV$_2$O$_4$ compared with that in MnV$_2$O$_4$.}
  \textbf{ (a)} Total energy versus angle and associated SIA (meV)
  of V$^{3+}$ in ambient (circle) and 10 GPa (square) pressure for bulk CoV$_2$O$_4$ and MnV$_2$O$_4$ (diamond).
  \textbf{ (b)} NC spin configurations of V$^{3+}$ and Co$^{2+}$/Mn$^{2+}$ spins
pointing along local [111] and global [001] directions, respectively.
  The round bold (dotted) arrows close to V spins depict the rotational flexibility in \co (MnV$_2$O$_4$).
 \textbf{ (c)} SIA of Co$^{2+}$ in bulk CoV$_2$O$_4$ under 10 Gpa compared to SIA of Mn$^{2+}$ in bulk MnV$_2$O$_4$.
  \textbf{ (d)} Orbital occupation configuration of Mn$^{2+}$ ($d^5$) and Co$^{2+}$ ($d^7$).
  \textbf{ (e)} $R_{V-V}$ in CoV$_2$O$_4$ and MnV$_2$O$_4$ are compared with $R_{V-V}$ in other vanadates from Ref.~\cite{Canosa}.
}
\end{figure}

First-principles calculations were used to explore the microscopic origin for the complex NC state in cubic CoV$_2$O$_4$.
As shown in Fig.~\ref{D}(a) and (c), the major magnetic anisotropy appears on the V$^{3+}$ site
with a magnitude two or three orders larger than for the $A$-site (Co/Mn).
Although the V$^{3+}$ ions are surrounded by similar octahedra in both \co and \mn,
the local [111] single-ion anisotropy (SIA) of V$^{3+}$
is significantly reduced in \co ($-1.2$ meV) compared to that in \mn ($-4.8$ meV)
due to the melting of OO by the pressure-induced itinerancy in CoV$_2$O$_4$.
Calculated by DFT, the SIA on the V sites totally
disappears with an external pressure around 10 Gpa in \co .

While the AFM V-V interaction in a pyrochlore lattice with local [111] SIA favors the all-in/all-out (AI/AO) spin structure,
the disappearance of SIA fosters strong magnetic frustration \cite{Ising}.
In \mn , that frustration was suppressed by OO.
However, the frustration reappears in \co due to the melting of the OO and the suppression of the
easy-axis anisotropy by itinerancy, as shown in Fig.~\ref{D}(a), (b).
The recovered frustration may be responsible for the macroscopic spin-glass behavior \cite{Kiswandhi}
below $T_{NC}$.

\begin{figure}
\includegraphics[width=0.91\textwidth]{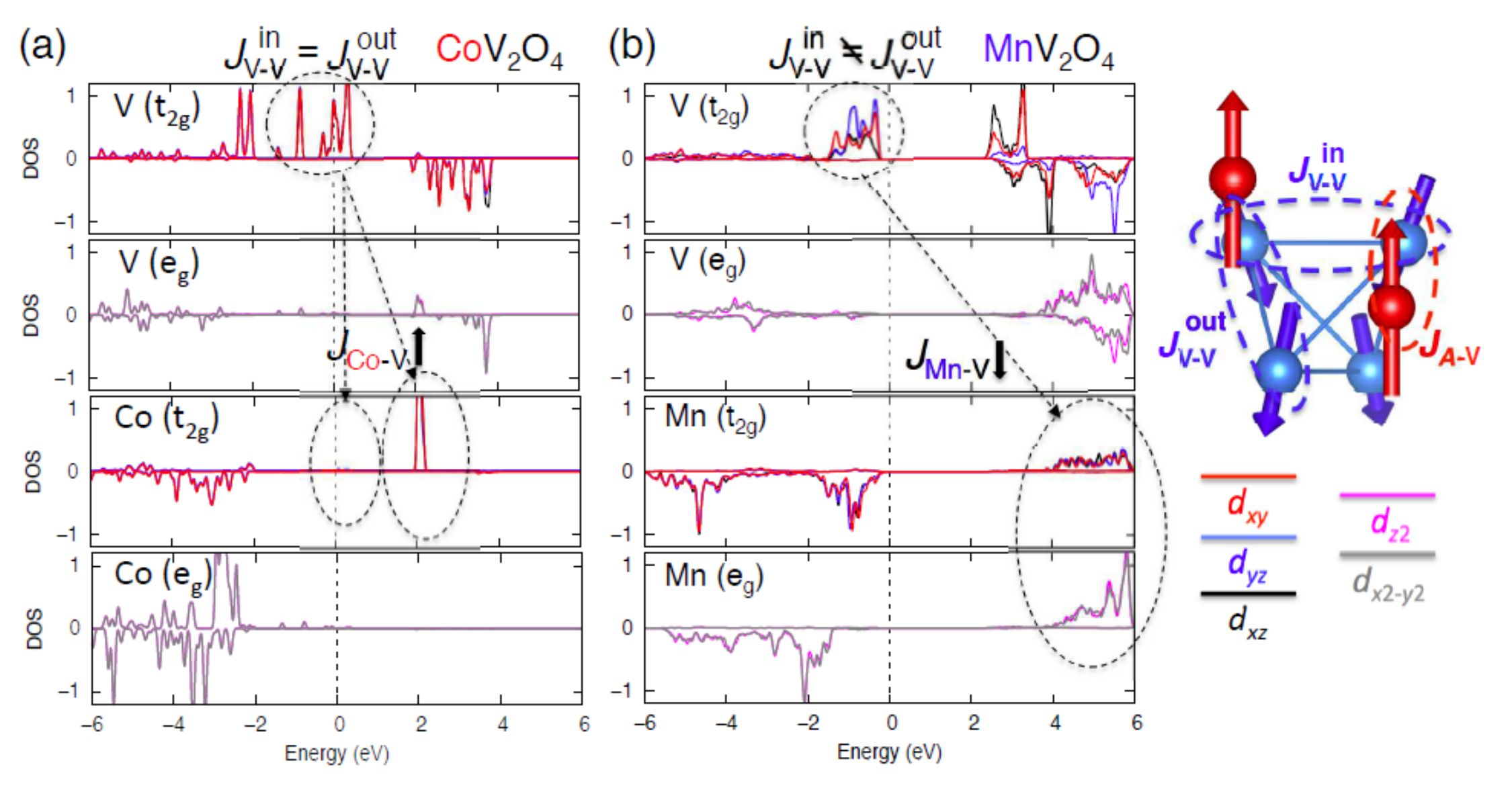}\\
\caption{\label{dos}
\textbf{ Origin of the enhanced magnetic ordering temperature in \co.}
Projected density-of-states of CoV$_2$O$_4$ \textbf{ (a)} compared to  MnV$_2$O$_4$ \textbf{ (b)} in unit of eV$^{-1}$.
Dotted arrows denote the energy differences, $\Delta$
between V and Co/Mn for possible AFM super-exchange ($J_{A-{\rm V}}$ $\sim$ -$t^2$/$\Delta$). $t$ is the hopping parameter between orbitals.}
\end{figure}

The SIA of $A$-site (Fig.~\ref{D}c), ($A$=Co, Mn) is quite negligible compared to the SIA of V$^{3+}$.
While Mn$^{2+}$ has a weak easy-plane axis because of the compressed tetragonal structure (c/a$<$1),
Co does not exhibit anisotropy because of the isotropic cubic structure.
The SIA of Co$^{2+}$ is much less dependent on pressure than that of V$^{3+}$ since Co$^{2+}$ electronic states lie
significantly below the Fermi energy ($\epsilon_F$) and are thereby electronically encapsulated, as shown in Fig.~\ref{dos}(a).
Only V$^{3+}$ states cross $\epsilon_F$.
Therefore the pressure-induced itinerancy will only affect the spins on V$^{3+}$ sites.

\vspace{4mm}
\noindent{\bf Enhanced exchange couplings}

\begin{figure}
\includegraphics[width=0.99\textwidth]{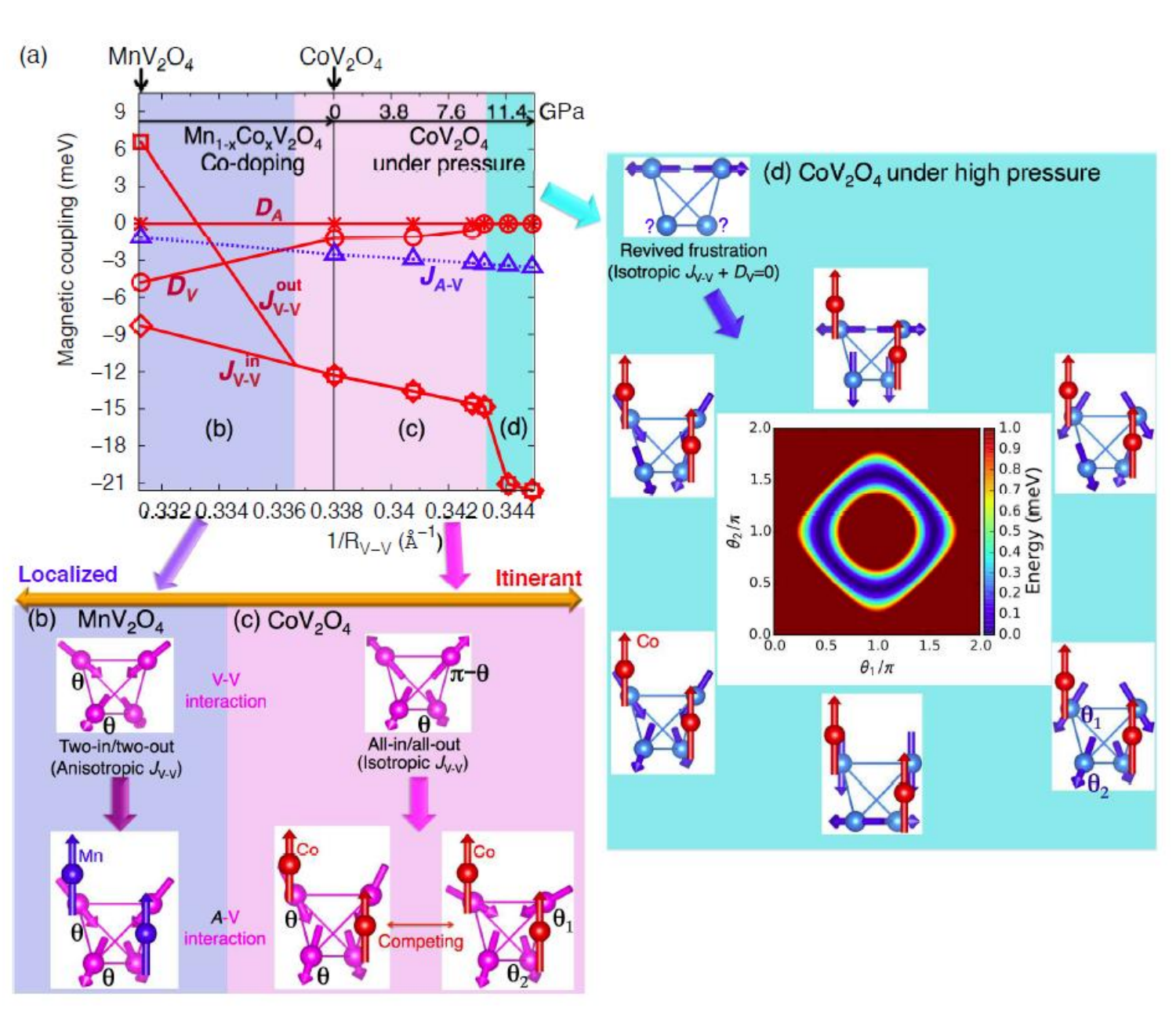}\\
\caption{\label{phase}
\textbf{ Evolution of magnetic couplings and competing ground states driven by Co-doping and pressure}
\textbf{ (a)} Change of all magnetic interactions with Co doping and external pressure (Gpa) calculated
  by LSDA+$U$ for the ground states at zero temperature.
  Points represent DFT results and the connecting lines are a guide for eye.
  $J_{\rm V-V}^{\rm in}$ and $J_{\rm V-V}^{\rm out}$ are expected to be degenerate at $x$=0.8 (cubic) in Mn$_{1-x}$Co$_x$V$_2$O$_4$
  from experiments \cite{Kiswandhi,Jie}.
 Bold (dotted) lines represent the exchange ($J$) and SIA ($D$) interactions.
\textbf{ (b)}  Origin of TI/TO state in Mn-rich region.
\textbf{ (c)} Competition of the isosymmetric one-angle TI/TO state with the two-angle state that evolves from the
AI/AO state. \textbf{ (d)} Spin glass phase appears with the disappearance of SIA.}
\end{figure}

As shown in Supplement, the Bragg peaks
do not split or broaden with decreasing temperature below 100 K,
indicating the absence of a strutural transition.
In agreement with this measurement, DFT calculations confirm the structural isotropy ($c/a=1$) of \co.
As shown in Fig.~\ref{dos}, the $t_{2g}$ ($d_{xy}$=$d_{yz}$=$d_{xz}$) and $e_g$ ($d_{z^2}$=$d_{x^2-y^2}$)
electronic levels become equally occupied and degenerate
in cubic \co . The structural and electronic isotropies
also produce the same exchange interactions $J^{\rm in}_{\rm V-V}$=$J^{\rm out}_{\rm V-V}$=$-12$ meV
between all spins on the tetrahedron as calculated from first principles, Fig.~\ref{phase}.
These coupled structural, electronic, and magnetic isotropies
foster frustration and the observed NC phase in Fig.~\ref{Bragg}.

Comparing the densities-of-states of \co and \mn reveals the origin of the enhanced magnetic ordering temperatures in \co .
The large energy difference ($\sim$ 5 eV) between the occupied V and Mn $d$ states weakens the
exchange between Mn and V. By filling the $e_g$ minority spin levels as indicated in Fig.~\ref{D}(d) and~\ref{dos}(a),
Co significantly lowers the $t_{2g}$ unoccupied energy level
and enhances the exchange interaction between Co and V.
DFT calculations reveal that the magnitude of the AFM $J_{AB}$ is twice as large in CoV$_2$O$_4$ (-2.5 meV) as in \mn (-1.2 meV).
As reflected by the neutron-scattering measurements in Fig.~\ref{Bragg},
the enhanced $J_{A-{\rm V}}$ causes $T_{CL}$ to more than double in \co (150 K) compared to \mn (53K).

Strikingly, the induced itinerancy also increases the NC ordering temperature even without OO in CoV$_2$O$_4$.
As shown in Fig.~\ref{Bragg}, $T_{NC}$ significantly increases in \co (75K) compared to \mn (57K).
Although it exhibits the higher NC ordering temperature, \co also exhibits glassy behavior \cite{Kiswandhi,Huang}.
While the reduced SIA and induced isotropies foster frustration \cite{Ising}, the enhanced exchange interaction
relieves the frustration and enhances the ordering temperatures.
In the series Mn$_{1-x}$Co$_x$V$_2$O$_4$, the spin-wave gap ($\sim$2meV)
remains relatively unchanged with Co-doping ($x$) \cite{Jie} despite the enhanced magnetic ordering temperatures proportional to $J_{\rm A-V}$.
Since the spin-wave gap is proportional to $\sqrt{D_{\rm V}\!\times\! J_{\rm A-V}}$,
the increase in $\vert J_{\rm A-V}\vert $ is compensated
by the reduction in the anisotropy $D_{\rm V}$ in \co.
By enhancing both competing effects (itinerancy-driven isotropies with reduced SIA and strengthened exchange), Co doping
can foster various novel states in CoV$_2$O$_4$.

\vspace{4mm}
\noindent{\bf Novel phases induced by frustration}

The comparison between \co and \mn in Fig.~\ref{phase} reveals the origin of the NC states in \co .
The key handle to tune the magnetic couplings is the distance between the V atoms ($R_{\rm V-V}$ along the $x$-axis) controlled by
chemical doping and external pressure.
In \mn, the OO of the V ions relieves the magnetic frustration of the pyrochlore lattice
and stabilizes the TI/TO NC spin state.
The AFM Mn-V interactions increase the canting angle while maintaining this TI/TO state (region {\bf b}).
By introducing itinerancy, Co doping promotes isotropic V-V interactions and
favors the AI/AO spin state.
Within the tetrahedron network, the AI/AO state has two distinct canting angles $\theta$ and $\pi$-$\theta$
compared to the one canting angle $\theta$ of the TI/TO state.
Guided by the DFT parameters for CoV$_2$O$_4$, our model calculation
indicates that the new two-angle state
based on the AI/AO state lies within 0.1 meV/unit-cell of the TI/TO ground state.
The isotropic exchange ($J_{\rm V-V}^{\rm in}=J_{\rm V-V}^{\rm out}$)
fosters a new two-angle AI/AO structure that can be stabilized by a magnetic field.

External pressure may also increase the degree of frustration.
For high external pressure \cite{Kismarahardja,Kiswandhi} $\sim$10 GPa,
the enhanced itinerancy fully suppresses the local SIA ($D_{\rm V}\sim$0) of V
as in Fig.~\ref{phase}(a) and revives the magnetic frustration of the pyrochlore lattice.
Although AFM exchange between the Co and V sites then induces the observed isosymmetric TI/TO spin structure,
the frustration fostered by itinerancy and the alternative states that compete with the TI/TO ground state
are responsible for the measured magnetic anomalies \cite{Kismarahardja}
and spin-glass behavior \cite{Kiswandhi}.

As shown in Figs.~\ref{D}(a) and \ref{phase}(a), high pressure may stabilize a continuum of degenerate states
where the two angles ($\theta_{1}$, $\theta_{2}$) rotate without energy cost due to the absence of SIA,
as obtained in the Appendix and shown in the
energy landscape of Fig.~\ref{phase}(d).
This degeneracy can induce spin-glass or spin-liquid-like behavior.
Since neutron scattering is limited to relatively low pressures,
these novel states should be studied with synchrotron magnetic X-ray scattering.

\vspace{4mm}
\noindent{\bf Magnetic field measurement}

\begin{figure}
\includegraphics[width=0.43\textwidth]{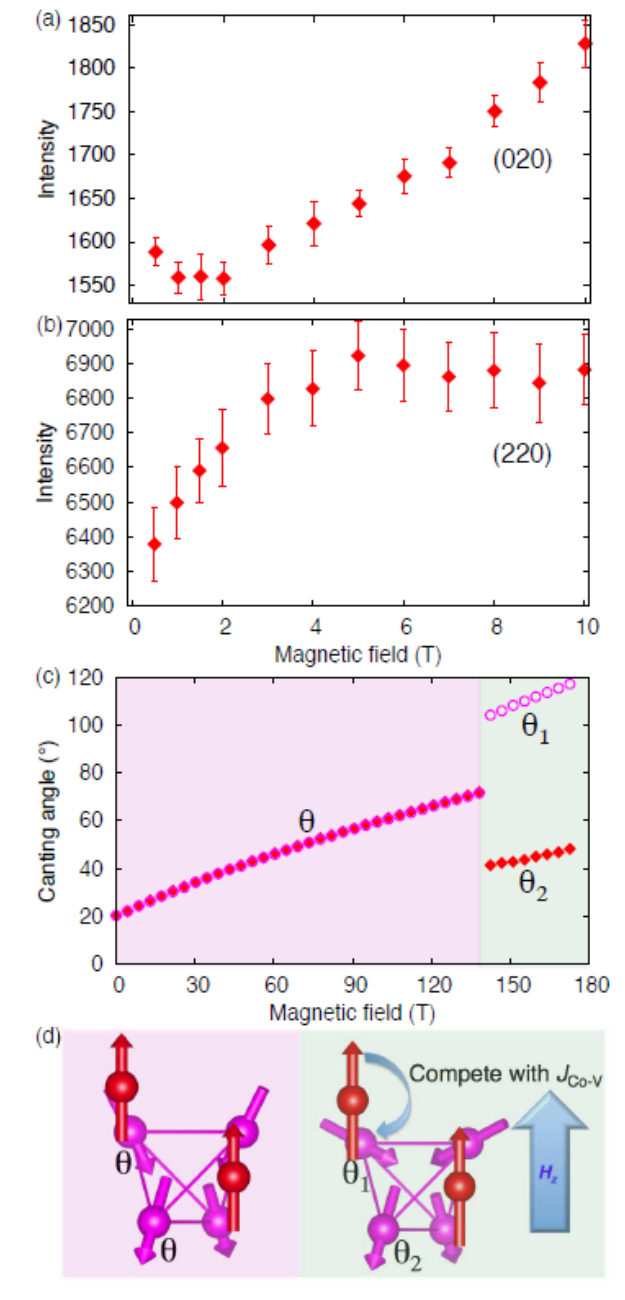}\\
\caption{\label{predict}
\textbf{ Prediction of novel phases driven by magnetic field in Mn$_{0.2}$Co$_{0.8}$V$_2$O$_4$.}
\textbf{ (a)} First-order phase transition from TI/TO to AI/AO-derived state
with magnetic field calculated by spin models using DFT parameters.
Inset is comparison with neutron scattering measurements (square) up to 10 T.
\textbf{ (b)} One-angle state based on TI/TO (left) and two-angle state based on AI/AO (right). The latter is driven
by the weakened $J_{\rm Co-V}$ in a magnetic field.}
\end{figure}

Fig.~\ref{phase}(a) provides a guide to uncover the novel states produced by the regenerated frustration.
Although all other magnetic couplings (isotropic $J_{\rm V-V}$, reduced $D_V$) foster frustration (bold red line),
the remnant AFM interaction $J_{\rm Co-V}$ (dotted blue line in Fig~\ref{phase}(a)) still relieves frustration.
An external magnetic field ($\vec{H}=H \hat{z}$) can help restore frustration
by weakening $J_{\rm Co-V}$, thereby inducing the novel two-angle state of \co, as shown in Fig.~\ref{predict}.

To check the effect of a magnetic field on CoV$_2$O$_4$, we carried out further elastic neutron-scattering measurements
on the Co-rich single-crystal spinel Co$_{0.8}$Mn$_{0.2}$V$_2$O$_4$,
which preserves the cubic structural and magnetic isotropies as in Fig.~\ref{phase}(a)
but exhibits stronger scattering intensity than \co due to the larger size of the single crystal.
The V$^{3+}$ AFM components in the $ab$-plane increase with the magnetic field ($H_z >$ 3 T),
as indicated by the increased intensity of (020) (see Fig.~\ref{predict}(a)).
At $H_z<$ 3 T, the increased intensity of (220)
reflects the reorientation of the magnetic domains; at $H_z >$ 3 T
the (220) intensity is saturated, indicating that all magnetic domains
are fully oriented and that the FIM components are constant.
This is consistent with magnetization measurements
on a polycrystalline sample, which provide a saturation field of $\sim$2 T \cite{Huang}.
Although our measurements can not disentangle the magnetic components along [001] ($M_z^{\rm Co}$ and $M_z^{\rm V}$),
we do not expect $\vert M_{z}^{\rm V}\vert $ to decrease with field above 3 T because that would require $M_z^{\rm Co}$
to further increase. So we can safely assume that both $M_{z}^{\rm V}$ and $M_{z}^{\rm Co}$ are constant above 3 T.
Since the AFM components of V$^{3+}$ in the $ab$-plane ($M_{ab}^{\rm V}$) continue to grow above 3 T,
the canting angle ($\theta={\rm tan}^{-1}[M_{ab}^{\rm V}/M_{z}^{\rm V}]$) of the V$^{3+}$ spins must
increase with the magnetic field along [001].

Using the spin model (Eq.~\ref{ModelHamiltonian}) combined with DFT parameters (Fig~\ref{phase}(a)),
we confirm the increase in the canting angle with magnetic field in Fig.~\ref{predict}(c).
The two-angle AI/AO state has an energy within 0.1meV/unit-cell of the
the one-angle TI/TO ground state in the Co-rich region.
We predict that
this new state is stabilized by a large magnetic field of about 140 T, as shown in Fig.~\ref{predict}(c),(d).
Although only the one-angle TI/TO state was previously reported
in vanadate compounds ($A$V$_2$O$_4$, $A$=Zn, Mn, Fe),
various competing states appear in CoV$_2$O$_4$ due to frustration.
It is likely that those states can be revealed by a magnetic field or pressure.

Of course, the critical magnetic field ($H_z$ = 140 T) is too large for neutron scattering measurements.
However, the first-order phase transition from the one-angle to the two-angle state
may be captured by magnetic susceptibility measurements.
Moreover, various methods can be employed to reduce the critical field.
Since external pressure suppresses SIA and revives frustration as discussed in the previous section,
pressure may also reduce the critical magnetic field.
Contrary to the usual expectation, a magnetic field
may strengthen frustration and noncollinearity
in \co by weakening the only exchange coupling ($J_{\rm Co-V}$) that hampers frustration.

\vspace{4 mm}
\noindent{\bf \large Discussion}

It is natural to wonder if cations significantly smaller than Co$^{2+}$ such as Be$^{2+}$ can be substituted on the $A$-site
to induce even more itinerancy and consequent frustration. However, a non-magnetic $A$-site cannot support localized V-spins
so the system would become paramagnetic~\cite{Canosa}.  Because strong magnetic interactions between the $A$ and $B$ sites is required to
maintain the localized V spins, Co is the only candidate $A$-site cation to support localized spins with enhanced
$J_{\rm Co-V}$ while inducing itinerancy on the $B$ site.

Compared to other vanadates ($A$V$_2$O$_4$), the frustration in magnetically and structurally isotropic \co
explains its NC and macroscopic spin-glass properties.
Since the AFM interaction between Co and V is the only factor
that relieves the magnetic frustration, weakening the AFM interaction by a magnetic field or
further reducing the SIA by external pressure can rekindle the frustration and reveal alternative states.
Among spinel vanadates, CoV$_2$O$_4$ is uniquely located at the crossover between localized and itinerant behavior.  Consequently,
many exotic properties and new phases can be produced by restoring the frustration of the pyrochlore lattice.

\vspace{4 mm}
\noindent{\bf \large Method}

\noindent{\bf Sample preparation}

Single crystals of CoV$_{2}$O$_{4}$, Co$_{0.8}$Mn$_{0.2}$V$_{2}$O$_{4}$ and MnV$_{2}$O$_{4}$ were grown by the
traveling-solvent floating-zone (TSFZ) technique. The feed and seed
rods for the crystal growth were prepared by solid state reaction.
Appropriate mixtures of MnO, CoCO$_{3}$, and V$_{2}$O$_{3}$ were ground together
and pressed into 6-mm-diameter 60-mm rods under 400 atm hydrostatic
pressure, and then calcined in Ar at 1050
$^{\circ}$C for 15 hours.  The crystal growth was carried out in
argon in an IR-heated image furnace (NEC) equipped with two halogen
lamps and double ellipsoidal mirrors with feed and seed rods
rotating in opposite directs at 25 rpm during crystal growth at a
rate of 20mm/h.

\vspace{4 mm}
\noindent{\bf Neutron-scattering experiments}

Single-crystal neutron diffraction was performed to determine the crystal and magnetic structures
using the four-circle diffractometer (HB-3A) at the High Flux Isotope Reactor (HFIR)
of the Oak Ridge National Laboratory (ORNL). A neutron wavelength of 1.003\AA~
was used from a bent perfect Si-331 monochromator \cite{chak11}.
High magnetic field single-crystal neutron diffraction experiments were performed
on the cold neutron triple-axis spectrometer (CTAX) at HFIR, ORNL.
The incident neutron energy was selected as 5.0 meV by a PG (002) monochromator,
and the final neutron energy was also set as 5.0 meV by a PG (002) analyzer.
The horizontal collimation was guide-open-80$^\prime$-open.  Contamination from higher-order beams was removed using a cooled Be filter.
The scattering plane was set in the (H,K,0) plane
and the magnetic field was applied perpendicular to the scattering plane.
The nuclear and magnetic structures were refined with the program FULLPROF \cite{rod93}.
Due to the domain re-orientation effect, intensities of both (220) and (020) diffractions
increase sharply in small magnetic fields,
but the (220) diffraction is saturated above about 3 T.
The intensity of the (020) diffraction, corresponding to the magnetic component of V in the $ab$-plane,
inceases linearly with field.

\vspace{4 mm}
\noindent{\bf First-principles calculations}

First-principles calculations were performed using density-functional theory
within the local spin-density approximation with
a correction due to on-site Hubbard interaction (LSDA+$U$)
as implemented in the Vienna {\it ab initio} simulation package (VASP-5.3) \cite{Kresse}.
We used the Liechtenstein \cite{Lich} implementation with on-site Coulomb
interaction $U$ = 6.0 eV and on-site exchange interaction
$J_{\rm H}$ = 1.0 eV to treat the localized 3d electron states in Co, Mn, and V;
this choice of $U$ is close to that chosen in previous work on CoV$_2$O$_4$ \cite{Kaur}
and MnV$_2$O$_4$ \cite{Nanguneri,Sarkar}. The spin-orbit interaction was included.
The projector augmented wave (PAW) potentials \cite{PAW1,PAW2} explicitly include 13 valenced electrons for Mn ($3p^63d^54s^2$),
9 for Co ($3d^84s^1$), 13 for V ($3s^23p^63d^44s^1$), and 6 for oxygen ($2s^22p^4$).
The wave functions were expanded in a plane-wave basis with an energy cutoff of 500 eV.
To evaluate the on-site single-ion anisotropy (SIA) interaction $D$, only one cation of interest was kept
while the surrounding magnetic atoms
were replaced by neutral and isoelectronic Ca$^{2+}$ and Al$^{3+}$ cations for Co$^{2+}$/Mn$^{2+}$ and V$^{3+}$, respectively.
This is the same technique that
was successfully used for BiFeO$_3$ \cite{wein12}
and CaMn$_7$O$_{12}$ \cite{zhang13}.

\vspace{4 mm}
\noindent{\bf Microscopic spin model}

Spin states in spinels can be described by the following model Hamiltonian,
\begin{equation}
\label{ModelHamiltonian}
H=  -\frac{1}{2}  \sum \limits_{ i,j} J_{ij} \boldsymbol{S}_i  \cdot \boldsymbol{S}_j + \sum \limits_{i} D_i \left( \hat{u_i} \cdot \boldsymbol{S}_i \right)^2
\end{equation}
which contains six inequivalent
sublattices. Isotropic exchange constants $J_{\rm Co-V}$ describe nearest-neighbor interactions between the Co and V sites.
$J_{\rm Co-Co}$ and  $J_{\rm V-V}$ describe nearest-neighbor interactions between Co-sites and V-sites, respectively.
The easy-axis anisotropy is assumed to be zero for the Co-sites, while for the B-site spins, the easy-axis anisotropy $D_{\rm V}$
is along the local $<$111$>$ direction.
The azimuthal directions of each vanadium spin
is constrained, but the canting angle $\theta_i$, described in Fig. \ref{phase},
is allowed to
vary between 0 and $2\pi$. Since $\theta_i$ may have a unique value in adjacent planes, both the two-in-two-out and all-in-all-out
configurations are possible. These angles are equal to the polar angle when $\theta_i$ is between 0 and $\pi$,
while the polar angles equals $2\pi-\theta_i$ and the azimuthal angle changes by $\pi$ when $\theta_i$ is greater than $\pi$.

The ground state spin configuration was found by minimizing the classical energy for a given set of parameters.
To avoid local minima, this was accomplished by calculating the classical energy on a grid with $\theta_i = 0$ to $2\pi$ and
finding the two angles with the lowest energy. This process was repeated for values of the external magnetic field ranging
from 0 to 173 T.

The inelastic neutron cross section for undamped spin waves was calculated using the
$1/S$ formalism outlined in Ref.~[\onlinecite{Haraldsen2009}] and the qppendices of Ref.~[\onlinecite{Fishman2013}].
For direct comparison with experimental intensities, the effects of the magnetic form factor
and the instrumental resolution were included in the calculation. The coefficients for Co$^{2+}$, and V$^{3+}$ are from
Ref.~[\onlinecite{neutrondatabooklet}]. The resolution function was approximated as a Gaussian in energy with a
full width at half-maximum of 1.5 meV. Effects from finite resolution in $\boldsymbol{Q}$ were not considered.

While DFT can provide guidance for the values of the isotropic exchange interactions,
LSDA+$U$ overestimates the
experimental moment (S$_{\rm V}$=0.23(7)) of \co measured by neutron scattering.
Our spin model uses the magnetic moment (S$_{\rm V}$=0.25), which
is within the experimental uncertainty. In addition, parameters calculated with DFT were adjusted to reproduce the
measured canting angle of \co ($\theta$ = 20.8 $\pm$ 1.7$^\circ$) in zero field.
Care was also taken to avoid a long-range spiral configuration~\cite{Tomiyasu04}
that was not observed in our neutron diffraction measurements.
The final set of
parameters used for \co are \mbox{$J_{\rm Co-Co} = 0.5\textrm{meV}$},
\mbox{$J_{\rm Co-V}=-2.5\textrm{meV}$}, \mbox{$J_{\rm V-V}= -11.0\textrm{meV}$},
\mbox{$D_{\rm V}=-2.7 \textrm{meV}$}, $S_{\rm Co} =$ 1.50 and $S_{\rm V} = 0.25$.

\vspace{4mm}
\noindent{\bf Canting angles of the new phase with ${\textbf {\emph{D}}_{\rm V}}$=0.0}

At high pressures, we take $D_{\rm V}$ = 0.0 in Eq.(\ref{ModelHamiltonian}).  Then,
the total energy per unit magnetic unit cell is

\begin{equation}
E\left( \alpha \right) = -12 J_{\rm Co-V} S_{\rm Co} S_{\rm V} \alpha - 4 J_{\rm V-V} S_{\rm V}^2 \alpha^2 + \textrm{constant}
\end{equation}
where $\alpha = \cos\theta_1 + \cos\theta_2$. The total energy is at a minimum if
\begin{equation}
\alpha = -\frac{3 J_{\rm Co-V} S_{\rm Co}}{2 J_{\rm V-V} S_{\rm V}}
\end{equation}
which limits the allowed combination of $\theta_1$ and $\theta_2$.
When $\theta_1 = \theta_2$ and $D_{\rm Co}$ = $D_{\rm V}$ = 0.0, this condition is equal to the expression
for $\theta$ in ref. \cite{Nanguneri}.

\vspace{4 mm}
\noindent{\bf  Acknowledgements}

The research at HFIR, ORNL, were sponsored by Department of Energy, Office of Sciences, Basic Energy Sciences, Materials Sciences and Engineering Division (J.H.L., S.O., R.F.) and Scientific User Facilities Division (J.M., S.E.H., M.M.). The research at UNIST was supported by Basic Science Research Program through the National Research Foundation of Korea (NRF) funded by the Ministry of Science, ICT $\&$ Future Planning (2.150639.01). J.M. thanks the support of the Ministry of Science and Technology of China (2016YFA0300500). S.E.H. acknowledges support by the Laboratory's Director's fund, ORNL. H.D.Z thanks the support from NSF with grant NSF-DMR-1350002. The authors acknowledge valuable discussions with G. MacDougall.

\vspace{4 mm}
\noindent{\bf Author contributions}

J.H.L. conceived the original idea and carried out first-principles calculations.
S.E.H. and R.S.F. performed spin-wave simulations.
J.M., H.C., T. H, and M.M. measured and analyzed neutron scattering Bragg peaks.
H.D.Z. synthesized the samples.
J.H.L., S.E.H., J.M., H.B.C., T.H., S.O., M.M, R.S.F. discussed the results.
J.H.L. and R.F. wrote the manuscript.

\vspace{4 mm}
\noindent{\bf Additional information}

{\bf Competing finalcial interests:} The authors delcare no competing financial interests.

\bibliographystyle{apsrev}

\newpage

\end{document}